\documentclass{PoS}

\usepackage{amsmath}
\usepackage{amsfonts}
\usepackage{amssymb}

\def\beq{\begin{equation}}
\def\eeq{\end{equation}}
\def\bsp#1\esp{\begin{split}#1\end{split}}
\def\beqa{\begin{eqnarray}}
\def\eeqa{\end{eqnarray}}

\def\sect#1{Sec.~{\ref{#1}}}

\def\Fig#1{Fig.~{\ref{#1}}}

\def\eqn#1{eq.~(\ref{#1})}
\def\Eqn#1{Equation~(\ref{#1})}
\def\eqns#1#2{eqs.~(\ref{#1}) and~(\ref{#2})}

\def\ord{{\cal O} }
\def\eps{\epsilon}

\title{Infrared singularities in the high-energy limit}

\ShortTitle{Infrared singularities at high-energy}

\author{\speaker{Lorenzo Magnea}\\
        Dipartimento di Fisica, Universit\`a di Torino, and INFN, Sezione di Torino\\
        E-mail: \email{magnea@to.infn.it}}

\author{Vittorio Del Duca\\
        INFN, Laboratori Nazionali di Frascati\\
        E-mail: \email{delduca@lnf.infn.it}}

\author{Claude Duhr\\
        Institute for Particle Physics Phenomenology, University of Durham\\
        E-mail: \email{claude.duhr@durham.ac.uk}}

\author{Einan Gardi\\
        Tait Institute, School of Physics and Astronomy, University of Edinburgh\\
        E-mail: \email{Einan.Gardi@ed.ac.uk}}

\author{Chris D. White\\
        School of Physics and Astronomy, Scottish University Physics Alliance,
        University of Glasgow\\
        E-mail: \email{Christopher.White@glasgow.ac.uk}}

\abstract{We use our current understanding of the all-order singularity 
structure of gauge theory amplitudes to probe their high-energy limit.
Our starting point is the dipole formula, a compact ansatz for the soft 
anomalous dimension matrix of massless multi-particle amplitudes.
In the high-energy limit, we find a simple and general expression for the
infrared factor generating all soft and collinear singularities of the amplitude,
which is valid to leading power in $|t|/s$ and to all logarithmic orders. This
leads to a direct and general proof of leading-logarithmic Reggeization for 
infrared divergent contributions to the amplitude. Furthermore, we can 
prove explicitly that the simplest form of Reggeization, based on the absence 
of Regge cuts in the complex angular momentum plane, breaks down at 
the NNLL level. Finally, we note that the known features of the high-energy 
limit can be used to constrain possible corrections to the dipole formula,
starting at the three-loop order.}

\FullConference{Loops and Legs in Quantum Field Theory - 11th DESY Workshop on Elementary Particle Physics,\\
		April 15-20, 2012\\
		Wernigerode, Germany}

\begin{document}

\section{The high-energy limit of gauge amplitudes}
\label{high}

The high-energy limit of scattering amplitudes, defined as the limit in which the 
center-of-mass energy of the colliding particles, $\sqrt{s}$, is much larger than
other relevant kinematic invariants, has been an interesting and lively subject
of studies for several decades~\cite{ELP,Collins,Forshaw:1997dc}. In renormalizable 
field theories, the high-energy limit is characterized by the presence of large 
logarithms, as is always the case for processes with two or more disparate 
energy scales. Such logarithms are of practical relevance, since they can spoil 
the convergence of perturbation theory, and they may need to be resummed; on 
the other hand, they are also theoretically interesting, since they are tied to the
infrared, semiclassical regime: this means that they can be computed in 
principle to all orders in perturbation theory, providing a useful handle on the
non-perturbative structure of the theory.

Studies of the high-energy limit began with non-relativistic potential models,
and the first powerful tool that was brought to bear was the analytic continuation
of amplitudes in the complex angular momentum plane~\cite{Regge}. The 
starting point for these studies is the well-known expansion of scattering
amplitudes in partial waves. In the case of a four-point amplitude, one writes
\beq
A(s,t) \, = \, 16 \pi \, \sum_{l = 0}^\infty \left( 2 l + 1 \right) \, a_l (s) \, P_l \left( 
\cos \theta \right) \, ,
\label{prtwa}
\eeq
where $P_l$ are Legendre polynomials, $\theta$ is the center-of-mass scattering 
angle,  in terms of which $t = - s (1 - \cos \theta)/2$, and $a_l(s)$ are the $s$-channel 
partial wave amplitudes. Using the causality and analyticity of the $S$ matrix as
fundamental postulates, one can analytically continue  the amplitude from the
physical region ($s>0$ and $t<0$) to the crossed $t$-channel, and construct
integral representations of the crossed-channel partial-wave amplitudes
$\tilde{a}_l(t)$. One finds then that the analytic properties of $\tilde{a}_l(t)$,
as functions of the angular momentum variable $l$, have powerful direct implications
on the high-energy behavior of the original amplitude $A(s,t)$. For example,
assuming that the only singularities of  $\tilde{a}_l(t)$ in the angular momentum 
plane are isolated poles, one finds that the large-$s$ asymptotic behavior of the 
amplitude is given by a simple power law,
\beq
  \tilde{a}_l (t) \sim \frac{1}{l - \alpha(t)} \, \quad \longrightarrow \quad
  A (s, t) \, \, \xrightarrow{\frac{|t|}{s}
  \rightarrow 0} \, \,  f(t) \, s^{\alpha(t)} \, ,
\label{Reggepo}
\eeq
where $\alpha(t)$ is the Regge trajectory.

This general result follows just from physically motivated postulates about 
the $S$ matrix, without reference to an underlying lagrangian field theory, 
and without resorting to a perturbative expansion. When one considers a
specific field theory model, and uses the tools of perturbation theory, one 
may verify that the structure schematically described by \eqn{Reggepo}
does indeed emerge. In a gauge theory, the high-energy limit is dominated
at tree level by the $t$-channel exchange of massless gauge bosons, and 
it is possible to show, at least at Leading Logarithmic (LL) accuracy~\cite{BFL}, 
that virtual corrections dress the $t$-channel gauge boson propagator 
according to
\beq
  \frac{1}{t} \, \longrightarrow \, \frac{1}{t} \, 
  \left( \frac{s}{- t} \right)^{\alpha(t)} \, ,
\label{propregg}
\eeq
where the Regge trajectory $\alpha(t)$ is now expressed as a perturbative expansion. 
A typical example of this structure is the resummed expression for the four-gluon 
amplitude in QCD, which can be written as
\beq
  M^{g g \rightarrow g g}_{a_1a_2 a_3 a_4} (s,t) 
  \, = \, 2 \, g_s^2 \, \frac{s}{t} \,
  \bigg[ (T^b)_{a_1a_3} C_{\lambda_1\lambda_3}(k_1, k_3) 
  \bigg] \, \left( \frac{s}{- t} \right)^{\alpha(t)} \,
  \bigg[ (T_b)_{a_2 a_4} C_{\lambda_2\lambda_4}(k_2, k_4) 
  \bigg] \, ,
\label{Mgg}
\eeq
where we label momenta so that $s = (k_1 + k_2)^2$ and $t = (k_1 - k_3)^2$.
The factorized form of \eqn{Mgg}, which has been shown to hold at NLL 
accuracy~\cite{Fadin:2006bj} for the real part of the amplitude, is consistent 
with \eqn{Reggepo}, and thus with the assumptions that the only singularities 
in the complex angular momentum plane should be isolated poles. 
The functions $C_{\lambda_i \lambda_j}(k_i, k_j)$ are the gluon impact 
factors, and are universal at least to the stated logarithmic accuracy: 
it is expected that one may use the same impact factor in other high-energy 
processes involving gluons as well as other energetic particles, for example 
quarks. In perturbative QCD, of course, virtual corrections to the 
amplitude are infrared divergent, and these divergences will appear in both
the impact factors and the Regge trajectory. To organize these divergences,
it is useful employ dimensional regularization (with $d = 4 - 2 \epsilon$ and 
$\epsilon < 0$), and express the trajectory as a perturbative series in powers
of the $d$-dimensional running coupling. One writes then
\beq
  \alpha(t) \, = \, \frac{\alpha_s (- t, \epsilon)}{4 \pi} 
  \, \, \alpha^{(1)} +
  \left( \frac{\alpha_s (- t, \epsilon)}{4 \pi} \right)^2 
  \, \alpha^{(2)} + \, \ord \left( \alpha_s^3 \right) \, ,
\label{alphb}
\eeq
where the $d$-dimensional running coupling~\cite{Magnea:1990zb} is given by
\beq
  \alpha_s (- t, \epsilon) = \left( \frac{\mu^2}{- t} 
  \right)^\epsilon \, \alpha_s (\mu^2) 
  + \, \ord \left( \alpha_s^2 \right) \, .
\label{rescal}
\eeq 
The first two perturbative coefficients $\alpha^{(1)}$ and $\alpha^{(2)}$ are 
known~\cite{Kuraev:1977fs,Fadin:1996tb,DelDuca:2001gu,
Sotiropoulos:1993rd,Korchemskaya:1996je}, and can be written as
\beq
  \alpha^{(1)} \, = \, C_A \, 
  \frac{\widehat{\gamma}_K^{(1)}}{\epsilon} \, ,
  \qquad
  \alpha^{(2)} \, = \, C_A \left[ - \frac{b_0}{\eps^2} +
  \widehat{\gamma}_K^{(2)} \,
  \frac{2}{\eps} + C_A \left( \frac{404}{27} - 2 \zeta_3 \right) 
  + n_f \left(- \frac{56}{27} \right) \right] \, .
\label{eq:2loop}
\eeq
where $b_0 = (11 C_A - 2 n_f)/3$, and $\widehat{\gamma}_K^{(i)}$ are the
perturbative coefficients of the cusp anomalous dimension~\cite{Korchemsky:1987wg}, 
with the overall Casimir eigenvalue $C_A$ scaled out. 

\begin{figure}
\begin{center}
\includegraphics[width=.4\textwidth]{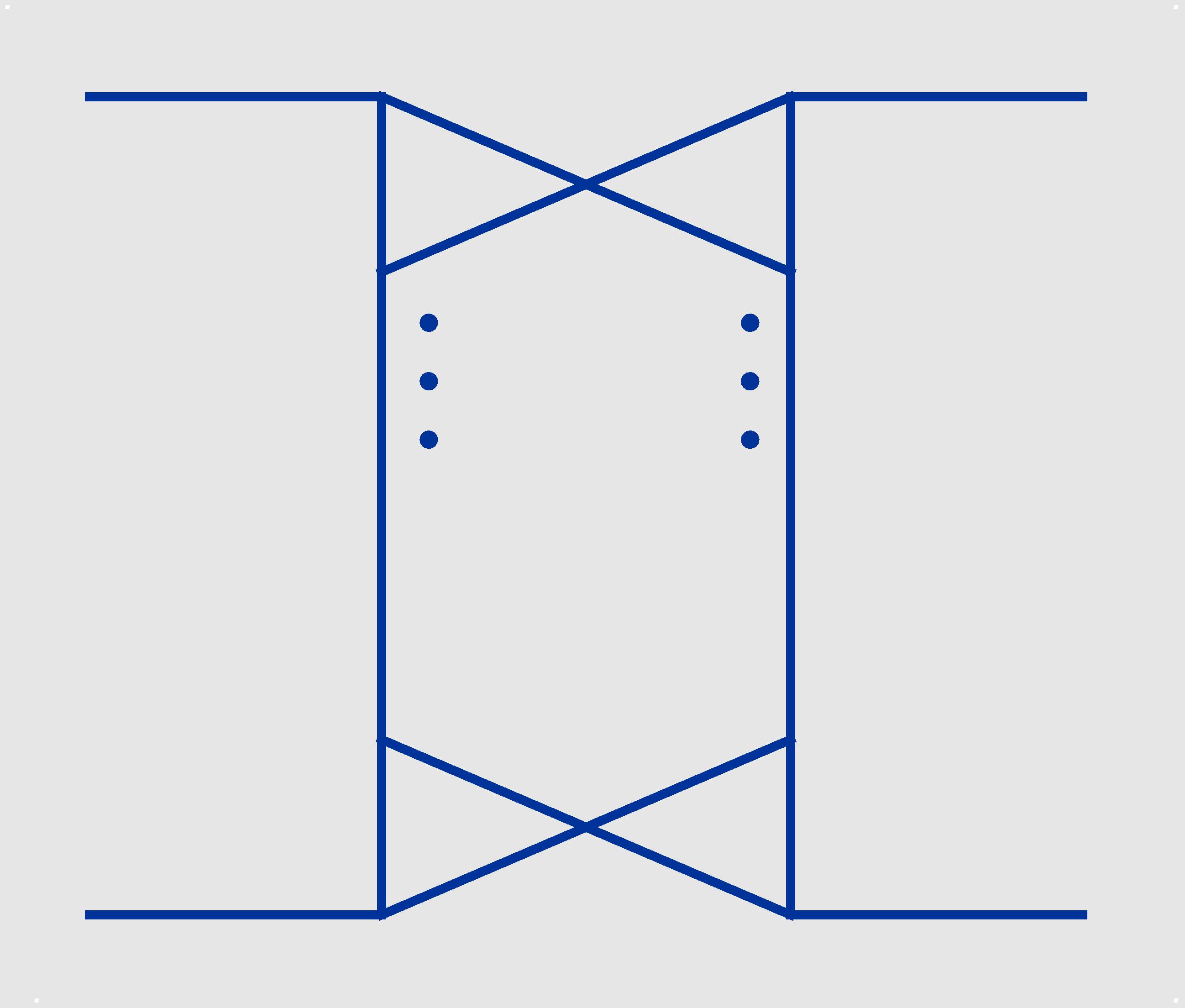}
\caption{The Mandelstam double-cross diagram.}
\label{Mandel}
\end{center}
\end{figure}

One must naturally wonder to what extent the factorization in \eqn{Mgg} can 
be trusted to higher logarithmic accuracy, a question which is intimately connected
to the underlying assumption that the only singularities in angular momentum
should be isolated poles. This question was studied in the early days of Regge 
theory, both on general grounds and in specific models (see, for example,
\cite{Collins}). In general, one may expect to find cuts as well as poles in
the complex angular momentum plane. It can be shown that, in a perturbative 
expansion, these `Regge' cuts require the $t$-channel exchange of at least two 
particles, and further require a non-planar diagrammatic structure. The simplest
class of diagrams with all the required features, which appears starting at three loops, 
is the `Mandelstam double-cross', portrayed in \Fig{Mandel}.
As we shall see, our approach to the high-energy limit will lead us to prove that
the simple factorized form of the amplitude given in \eqn{Mgg} must break down
at NNLL, and the first energy logarithm not correctly predicted by a naive 
Regge-pole-based factorization arises at three loops, from non-planar diagrams,
in agreement with the expectation that Regge cuts should play a role at precisely
that level of accuracy.

In the following, we will briefly describe our knowledge of the general structure
of infrared poles in multi-particle fixed-angle massless gauge theory amplitudes,
and we will discuss how this knowledge can be applied to the high-energy limit, summarizing the results of ~\cite{DelDuca:2011xm}.
We will show that infrared divergences naturally `Reggeize', for general $t$-channel
color exchanges, and we will prove that the simplest form of Reggeization breaks 
down at NNLL accuracy. Interestingly, the general results of Regge theory also
have implications on the structure of infrared poles, providing non-trivial constraints
on possible corrections to the simplest ansatz for the soft anomalous dimension,
the dipole formula~\cite{Gardi:2009qi,Becher:2009cu}.

\section{The infrared structure of gauge amplitudes}
\label{infra}

Let us consider a scattering amplitude in a massless gauge theory, involving
$n$ colored particles in arbitrary representations of the gauge group. Such an 
amplitude can be seen as a vector in the vector space spanned by all possible
color tensors connecting the various representations involved. One writes
\beq
  M_{a_1 \ldots a_n} \left(\frac{p_i}{\mu}, \alpha_s (\mu^2),
  \epsilon \right) \, = \,
  \sum_J  \, M_J \left( \frac{p_i}{\mu}, \alpha_s (\mu^2), 
  \epsilon \right) \, (c_J)_{a_1 \ldots a_n} \, ,
\label{ampcol}
\eeq
where the color tensors $c_J$ form a basis in the relevant vector space, $p_i$ 
are the particle momenta, and $a_i$ their color indices.

Such on-shell matrix elements are plagued by soft and collinear divergences order 
by order in perturbation theory. Studies performed over the space of several decades 
(see~\cite{Sterman:1995fz} for a review), and summarized in this context in 
Refs.~\cite{Gardi:2009qi,Dixon:2008gr}, have shown that infrared singularities
factorize. Denoting by $M$ the vector with components $M_J$, one can 
write~\cite{Gardi:2009qi,Becher:2009cu} the matrix equation
\beq
  M \left(\frac{p_i}{\mu}, \alpha_s (\mu^2),  \epsilon 
  \right) \, = \, Z \left(\frac{p_i}{\mu_f}, \alpha_s(\mu_f^2),  
  \epsilon \right) \, \, H \left(\frac{p_i}{\mu}, 
  \frac{\mu_f}{\mu}, \alpha_s(\mu^2), \epsilon \right) \, ,
\label{Mfac}
\eeq
where $H$ is a vector of matching coefficients, which are finite as $\epsilon \to 0$,
while all infrared singularities are collected in the matrix factor $Z$, and $\mu_f$
is a factorization scale. A key property of the infrared factor $Z$ is that it obeys a 
(matrix) renormalization group equation with a finite anomalous dimension 
$\Gamma$, which reads
\beq
  \mu \frac{d}{d \mu} \, Z \left(\frac{p_i}{\mu}, \alpha_s(\mu^2), \epsilon \right) 
  \, = \, - \,  Z \left(\frac{p_i}{\mu}, \alpha_s(\mu^2), \epsilon \right) \,
  \Gamma \left(\frac{p_i}{\mu}, \alpha_s(\mu^2) \right) \, .
\label{RG}
\eeq
This equation is easily, if formally, solved in exponential form, as
\beq
  Z \left(\frac{p_l}{\mu}, \alpha_s(\mu^2), \epsilon \right) \, = \,  
  P \exp \left[ \frac{1}{2} \int_0^{\mu^2} \frac{d \lambda^2}{\lambda^2} \, \,
  \Gamma \left(\frac{p_i}{\lambda}, \alpha_s(\lambda^2) \right) \right] \, ,
\label{RGsol}
\eeq
where we have used as a boundary condition the fact that the $d$-dimensional 
running coupling vanishes in the infrared for $d>4$, so that one can set
$Z(\mu = 0) = 1$. The anomalous dimension matrix $\Gamma$ is the centerpiece
dictating the infrared structure of multi-particle amplitudes. The key result of 
Refs.~\cite{Gardi:2009qi,Becher:2009cu} is that $\Gamma$ itself, in the case of 
massless particles, obeys a set of exact evolution equations which strongly constrain 
its kinematic dependence, tying it to the color structure of the process. The simplest
all-order solution to this set of equations is the dipole ansatz for the matrix $\Gamma$,
which reads
\beq
  \Gamma_{\rm dip}  \left(\frac{p_i}{\lambda}, \alpha_s(\lambda^2) \right) \, = \,
  \frac{1}{4} \, \widehat{\gamma}_K \left(\alpha_s (\lambda^2) \right) \,
  \sum_{(i,j)} \ln \left(\frac{- s_{ij}}{\lambda^2} 
  \right) {\bf T}_i \cdot {\bf T}_j \, - \, \sum_{i = 1}^L
  \gamma_{i} \left(\alpha_s (\lambda^2) \right) \, .
\label{sumodipoles}
\eeq
\Eqn{sumodipoles} must be understood as a color operator acting on the hard 
coefficients $H$ by means of the gluon insertion operators ${\bf T}_i$, which act 
exclusively on the color indices of the $i$-th particle, modifying the color structure 
according to the Feynman rules for single gluon emission~\cite{Bassetto:1984ik}. 
The solution is derived under the assumption that the cusp anomalous dimension 
corresponding to Wilson lines in color representation $R$, $\gamma_K^{(R)} 
(\alpha_s)$, is proportional, to all orders, to the quadratic Casimir eigenvalue 
$C_R$; one may then define the universal function $\widehat{\gamma}_K \left(
\alpha_s\right)$ according to $\gamma_K^{(R)} (\alpha_s) = C_R \widehat{
\gamma}_K \left(\alpha_s\right)$. The functions $\gamma_i (\alpha_s)$, on the 
other hand, carry no color structure, and are simply given by the anomalous 
dimensions of the fields corresponding to the the each hard external particle. \Eqn{sumodipoles} has several remarkable features, which cannot be reviewed 
in detail here. Let us simply note that $\Gamma_{\rm dip}$ contains only two-particle 
correlations, which mirrors the structure of one-loop corrections; it thus preserves 
the simplicity of single-gluon exchange. This implies vast cancellations in 
higher-order diagrammatic calculations, and means in particular that the color
structure $\Gamma_{\rm dip}$ is fixed at one loop, so that the path-ordering 
symbol in \eqn{RGsol} becomes superfluous. One also sees how the structure 
of double, soft-collinear poles is generated, even as one is simply solving a 
renormalization group equation, which would ordinarily generate only single 
poles: indeed, integrating the $d$-dimensional running coupling, accompanied 
by an extra logarithm  of the scale as in \eqn{sumodipoles}, generates precisely 
such poles.

Corrections to the dipole ansatz can arise from precisely two sources. First,
it is possible that the cusp anomalous dimension may cease to be proportional to 
the quadratic Casimir eigenvalue, starting at some loop order. This can in principle
happen starting at four loops, where quartic Casimirs can arise, however arguments
have been given in~\cite{Becher:2009cu,Vernazza} indicating that this is in fact not 
the case, so that the first correction of this kind should be confined to even higher 
orders. The only other possible corrections to \eqn{sumodipoles}, which can arise 
starting at three loops and for at least four external particles, must take the form
of functions of conformal invariant cross-ratios of external momenta, $\rho_{ijkl}
\equiv (p_i \cdot p_j p_k \cdot p_l)/(p_i \cdot p_k p_j \cdot p_j)$. One writes then 
in general
\beq
  \Gamma \left(\frac{p_i}{\lambda}, \alpha_s(\lambda^2) \right) \, = \,
  \Gamma_{\rm dip} \left(\frac{p_i}{\lambda}, \alpha_s(\lambda^2) \right) 
  \, + \, \Delta \left( \rho_{ijkl}, \alpha_s (\lambda^2) \right) \, .
\label{Delta}
\eeq
The function $\Delta$ has been studied in detail in~\cite{Dixon:2009ur,Vernazza}.
A number of constraints can be imposed on $\Delta$, but while they severely limit 
the available functional forms, they are not sufficient to prove that it vanishes, even 
at the three loop level. Whether such quadrupole corrections do indeed arise, starting 
at three loops, is a question that will probably need to be answered by a direct 
calculation.

\section{Infrared constraints on high-energy logarithms}
\label{infraco}

Studying the high-energy limit of the dipole formula is clearly of interest: indeed, 
the perturbative Regge trajectory is infrared divergent, and one may expect to be 
able to make general predictions, at least for divergent terms, to all orders in 
perturbation theory. In order to proceed, one must first note that the high-energy 
limit is formally outside the domain of applicability of the factorization theorem 
expressed by \eqn{Mfac}, which was derived for fixed-angle amplitudes, {\it i.e.} 
assuming that all kinematic invariants are of the same order, and all large with 
respect to the strong interaction scale $\Lambda$. This objection does not 
affect our arguments: we can start with a fixed-angle configuration, where the
factorized expression applies, and continuously deform the kinematics towards 
the high-energy limit, while staying in the perturbative regime; what happens is 
that energy logarithms, $\ln (s/|t|)$, become large, and they fail to properly factorize, 
but no new infrared poles are generated. We are thus in a position to make predictions, 
at least for the infrared divergent part of the various ingredients entering the 
Regge-factorized form of the amplitude.

Let us consider, for simplicity, the four-point amplitude (analogous considerations
apply to the multiparticle amplitudes in multi-Regge kinematics, as shown explicitly 
in~\cite{DelDuca:2011xm}). The key property of the infrared factor $Z$ in the Regge
limit $s \gg |t|$ is that, to leading power in $|t|/s$, but to all logarithmic accuracies,
it factorizes into the product of two factors, one of which carries the $s$ dependence
and the non-trivial color information, while the second one depends only on $t$
and is proportional to the identity in color space. We write this result as
\beq
  Z \left( \frac{p_i}{\mu}, \alpha_s (\mu^2), \epsilon \right)
  \, = \, \widetilde{Z} \left( \frac{s}{t}, \alpha_s (\mu^2), \epsilon 
  \right) \, Z_{\bf 1} \left( \frac{t}{\mu^2}, \alpha_s (\mu^2), 
  \epsilon \right) \, .
\label{Zfac}
\eeq
The color-singlet factor $Z_{\bf 1}$ contains collinear (as well as soft) divergences
to be associated with impact factors in a Regge-factorized expression. The central
piece of \eqn{Zfac} is the $s$-dependent matrix $\widetilde{Z}$, which can be written
in a very general and elegant way as
\beq
  \widetilde{Z} \left( \frac{s}{t}, \alpha_s (\mu^2), \epsilon \right) 
  \, = \, \exp \left\{ K \Big(\alpha_s (\mu^2), \epsilon \Big)
  \Bigg[ \ln \left( \frac{s}{- t} \right) {\bf T}_t^2 + 
  {\rm i} \pi \, {\bf T}_s^2 \Bigg] \right\} \, .
  \label{Ztildedef}
\eeq
Here we have introduced notations for the color insertion operators 
corresponding to the color representations exchanged in the $t$ and $s$ 
channel~\cite{Dokshitzer:2005ig}, defining
\beq 
{\bf T}_s \equiv {\bf T}_1  +  {\bf T}_2 \quad , \quad  {\bf T}_t \equiv {\bf T}_1  
+  {\bf T}_3 \, ,
\label{TsTt}
\eeq
for the scattering process $1 + 2 \longrightarrow 3 + 4$. Furthermore, we have 
introduced the notation
\beq
K \Big(\alpha_s (\mu^2), \epsilon \Big) \, \equiv \,  
  - \frac14 \int_0^{\mu^2} \frac{d \lambda^2}{\lambda^2} \, 
  \widehat{\gamma}_K \left(\alpha_s(\lambda^2, \epsilon) 
  \right) \, . 
  \label{Kdef} 
\eeq
It is straightforward to show that \eqns{Zfac}{Ztildedef} imply LL Reggeization of
infrared poles in $t$-channel exchanges for the scattering of generic color 
representations. Indeed, at LL accuracy, one may neglect the $s$-channel phase
in the exponent of \eqn{Ztildedef}, which starts contributing at NLL. One can then 
write the full amplitude as
\beq
  \left. M \left( \frac{p_i}{\mu}, 
  \alpha_s(\mu^2), \epsilon \right) \right|_{LL} \, = \, 
  \exp \left\{ K \Big( \alpha_s (\mu^2), 
  \epsilon \Big) \, \ln \left(\frac{s}{- t} \right) \,
  {\bf T}_t^2 \right\} \, \, Z_{\bf 1} \, \, H \left( \frac{p_i}{\mu}, 
  \alpha_s(\mu^2), \epsilon \right) \, .
\label{Mggdef}
\eeq
Whenever the Born amplitude (which gives the leading-order contribution to $H$) is
dominated by $t$-channel exchanges at leading power in $|t|/s$, each such exchange 
is an eigenstate of the $t$-channel color operator ${\bf T}^2_t$, with an eigenvalue
given by the appropriate Casimir operator. For gluon-gluon scattering, for example,
the hard scattering is dominated in the high-energy limit by gluon exchange, and one 
may write
\beq
  {\bf T}_t^2 \, H^{gg\rightarrow gg} \, =
  \, C_A \, H^{gg\rightarrow gg}_t \, + \, \ord{\left(|t|/s \right)} \, .
\label{eigen}
\eeq
This automatically leads to Reggeization, yielding
\beq
  M^{gg\rightarrow gg}  \, = \,
  \left(\frac{s}{-t}\right)^{C_A \, K \left(\alpha_s (\mu^2),
  \epsilon \right)} Z_{\bf 1} \, H^{gg\rightarrow gg}_t \, ,
\label{Mgg3}
\eeq
and thus providing an all-order expression for the Regge trajectory, given by \eqn{Kdef}, reproducing early results~\cite{Korchemskaya:1996je}.

Since the factorization in \eqn{Ztildedef} is valid at leading power in $|t|/s$, we can 
in principle extend our predictions to all subleading energy logarithms. One way to do 
it is to use the Baker-Campbell-Haussdorff formula to express $\widetilde{Z}$ as a 
product of exponentials. One finds
\beqa
  && \hspace{-1mm} \widetilde{Z} \left( \frac{s}{t}, \alpha_s (\mu^2) , 
  \epsilon \right)  \, =  \, \left( \frac{s}{- t} \right)^{K \, \, {\bf T}_t^2} \, \, 
  \exp \left\{ {\rm i} \, \pi \, K \, \, {\bf T}_s^2 \right\} \,
  \exp \left\{ - \, {\rm i} \, \frac{\pi}{2}  \, K^2 \, 
  \ln \left(\frac{s}{- t} \right) \,  
  \Big[{\bf T}_t^2, {\bf T}_s^2 \Big] \right\} \label{Zcomm}
   \\ & & \hspace{-1mm} \times \, 
  \exp \left\{ \frac{K^3}{6} \left(- 2 \pi^2 \ln 
  \left(\frac{s}{- t}\right) 
  \Big[ {\bf T}_s^2, \big[{\bf T}_t^2, {\bf T}_s^2 \big] \Big] 
  \, + \,  {\rm i} \pi \, \ln^2 \left(\frac{s}{- t} \right) 
  \Big[{\bf T}_t^2, \big[{\bf T}_t^2, {\bf T}_s^2 
  \big] \Big] \right) \right\} \,\, \exp \left\{ {\cal O} \left( K^4 \right) \right\} \, .
  \nonumber
\eeqa.
Each exponential factor in \eqn{Zcomm} carries increasing powers of the 
function $K(\alpha_s, \epsilon) \sim \alpha_s/\epsilon$. One observes that, 
as expected, only the first factor contributes at LL. At NLL accuracy, one finds 
a series of commutators, which in general are not diagonal in the $t$-channel 
basis, and will thus break the naive form of Regge factorization expressed for 
example by \eqn{Mgg}. All these commutators however appear as phases, 
so that they will contribute only to the imaginary part of the amplitude: we see, 
on the other hand, that infrared poles for the real part of the amplitude Reggeize 
at NLL accuracy for arbitrary color exchanges in the $t$ channel, extending the 
results of~\cite{Fadin:2006bj}. At NNLL accuracy, we predict that the simple form 
of Regge factorization, which is based on the assumption that only Regge poles 
arise in the angular momentum complex plane, must break down also for the 
real part of the amplitude. The first color operator effecting this breakdown 
arises at ${\cal O} (\alpha_s^3)$, and it is given by
\beq
  E \left( \frac{s}{t}, \alpha_s, 
  \epsilon \right) \equiv - \, \frac{\pi^2}{3} \,
  {K^3 (\alpha_s, \epsilon)} \, \ln \left(\frac{s}{- t} \right) 
  \Big[{\bf T}_s^2, \big[{\bf T}_t^2, {\bf T}_s^2 \big] \Big] \, .
\label{NNLL}
\eeq
The most important thing to note concerning \eqn{NNLL} is that Regge-breaking
terms start arising at three loops, at NNLL level, and they involve non-planar
contributions to the amplitude, as implied by the presence of color commutators; 
furthermore, they arise from the squaring of the $s$-channel phase in \eqn{Ztildedef}.
All these features closely match what is expected from the presence of Regge cuts
in the angular momentum plane, as discussed in \sect{high}. A second important point 
is that possible corrections to the dipole formula arising at three loops (the function 
$\Delta$ in \eqn{Delta}) cannot rescue naive Regge factorization, since they contribute
terms of order $\alpha_s^3/\epsilon$ to the amplitude, whereas \eqn{NNLL} contains
terms of order $\alpha_s^3/\epsilon^3$. We also note that a trace of this Regge-breaking
mechanism must arise already at two loops in terms with no energy logarithms, which 
are generated  for example from the expansion of the second factor in \eqn{Zcomm}. 
Terms of precisely this form, proportional to $\alpha_s^2 \pi^2/\epsilon^2$, where indeed 
detected in~\cite{DelDuca:2001gu} by direct calculation. Finally, we note that double 
color commutators terms such as those generated by \eqn{NNLL} have recently
been shown to play a role~\cite{Forshaw:2012bi} in the breakdown of strict collinear 
factorization described in~\cite{Catani:2011st}: in that context, these terms arise 
from the contributions of Glauber gluons which survive in finite hadronic cross 
sections which are not fully inclusive, in the form of super-leading 
logarithms~\cite{Forshaw:2008cq}.

\section{High-energy constraints on infrared poles}
\label{HE}

We close by briefly illustrating how the known structure of the Regge limit can be 
used to impose further constraints on quadrupole corrections to the dipole formula,
which may arise starting at the three loop order. Considering again for simplicity
the four-point amplitude, and following Ref.~\cite{Dixon:2009ur}, we note that the
three relevant conformal cross ratio can be written as
\beq
  \rho_{1234}  \, =  \,  \left(\frac{s}{- t}\right)^2 \, {\rm e}^{- 2 {\rm i} \pi }
  \, ,  \qquad  
  \rho_{1342} \, = \, \left(\frac{- t}{s + t}\right)^2 \, ,  \qquad  
  \rho_{1423} \, = \, \left(\frac{s + t}{s}\right)^2 \, {\rm e}^{2{\rm i} \pi} \, , 
\label{L423_forward}
\eeq
where the appropriate phases have been kept, and the constraint $\rho_{1234} 
\rho_{1342} \rho_{1423} = 1$ is verified. Ref.~\cite{Dixon:2009ur}, after examining
the constraints imposed on the function $\Delta$ in \eqn{Delta} by soft-collinear
factorization, Bose symmetry and transcendentality, found a small set of 
functions that would still give allowed contributions. The functions considered 
in~\cite{Dixon:2009ur} have simple dependences on logarithms or polylogarithms
of the conformal cross ratios in \eqn{L423_forward}, and one can study their
high-energy limit. Defining $L \equiv \ln(s/|t|)$, and $L_{ijkl} \equiv \ln \rho_{ijkl}$,
one readily sees that, to leading power in $s/|t|$, the logarithms $L_{ijkl}$ become
\beq
  L_{1234}  \, = \, 2 ( L - {\rm i} \pi) \, , \qquad
  L_{1342}  \, = \, - 2 L \, , \qquad 
  L_{1423}  \, = \, 2 {\rm i} \pi \, .
\label{L423_forward_log}
\eeq
One can now examine what happens to the non-trivial examples of $\Delta$ that were discussed in~\cite{Dixon:2009ur}. The simplest case, involving only logarithms, is the function
\beqa
\label{Delta_case212}
  \hspace{-5mm}
  \Delta^{(212)} (\rho_{ijkl}, \alpha_s) \, = \, 
  \left( \frac{\alpha_s}{\pi} \right)^3 \,
  {\bf T}_1^{a} {\bf T}_2^{b} {\bf T}_3^{c} {\bf T}_4^{d} \,
  \bigg[ && \! \! f^{ade} f^{cbe} \,
  L_{1234}^2 \, \Big(L_{1423} \, L_{1342}^{2} \, + \, 
  L_{1423}^{2} \, L_{1342} \Big) \\
  + && \! \! f^{cae} f^{dbe} \, 
  L_{1423}^2 \, \Big(L_{1234} \, L_{1342}^{2} \, + \, 
  L_{1234}^{2} \, L_{1342} \Big) \nonumber \\
  + && \! \! f^{bae} f^{cde} \,
  L_{1342}^{2} \, \Big(L_{1423} \, L_{1234}^{2} \, + \,
  L_{1423}^{2} \, L_{1234} \Big) \bigg] \, .
  \nonumber
\eeqa
It is now immediate to verify that, in the high-energy limit
\beqa
  \Delta^{(212)} (\rho_{ijkl}, \alpha_s) & = & 
  \left(\frac{\alpha_s}{\pi}\right)^3 \, {\bf T}_1^{a} {\bf T}_2^{b} 
  {\bf T}_3^{c} {\bf T}_4^{d} \, \,
  32 \, {\rm i} \, \pi \Big[ \Big( - L^4 - {\rm i} \pi L^3
  - \pi^2 L^2 - {\rm i} \pi^3 L \Big) f^{ade}f^{cbe}
  \nonumber \\ && \qquad + \, \,
  \Big(2 {\rm i} \pi L^3 - 3 \pi^2 L^2 - {\rm i} \pi^3 L \Big) 
  f^{cae}f^{dbe} \Big] \, + {\cal O} \left( |t/s| \right)\, .
\label{Delta_4_large_s}
\eeqa
The known structure of the Regge limit forbids the appearance of such a function,
since it would generate super-leading high-energy logarithms of the form
$\alpha_s^n L^{n+1}$ at three loops and beyond. Studying the high-energy 
limit is sufficient to rule out, individually, all explicit examples of $\Delta$ that
were discussed in~\cite{Dixon:2009ur}. It is clear however that it may still be
possible to consider linear combinations of the same functions, constructed 
precisely in order to cancel the logarithms that happen to be constrained by 
Regge theory arguments. Indeed, a recent study~\cite{Vernazza} has identified
sets of functions that explicitly satisfy all known constraints, including those 
arising from the high-energy limit. The question of the existence of non-vanishing corrections to the dipole formula at three loops and beyond remains therefore
open.

\end{document}